\documentclass[aps,prl,onecolumn,preprint,groupedaddress]{revtex4}
\usepackage{graphicx,color,ulem}% Include figure files
\usepackage{dcolumn}% Align table columns on decimal point	
\usepackage{bm}% bold math
\usepackage{tabularx}
\usepackage{ulem}
\begin{document}

\title{Rejuvenation engineering in metallic glasses by complementary stress and structure modulation}% Force line breaks with \

\author{D.~\c Sopu$^{1,2}$}
\email[Corresponding Author: ]{daniel.sopu@oaew.ac.at}
\author{F~.Spieckermann$^3$}
\email[Corresponding Author: ]{florian.spieckermann@unileoben.ac.at}
\author{X.L. Bian$^4$}
\email[Corresponding Author: ]{bianxilei@shu.edu.cn}
\author{S. Fellner$^1$}
\author{J. Wright$^5$}
\author{M. Cordill$^1$}
\author{C. Gammer$^1$}
\author{G. Wang$^4$}
\author{ M.~Stoica$^5$}
\author{ J.~Eckert$^{1,3}$}
% Affiliations / Addresses (Add [1] after \address if there is only one affiliation.)
\address{%
$^{1}$Erich Schmid Institute of Materials Science, Austrian Academy of Sciences, Jahnstra{\ss}e 12, A-8700 Leoben, Austria\\
$^{2}$Institut f\"ur Materialwissenschaft, Fachgebiet Materialmodellierung, Technische Universit\"at Darmstadt, Otto-Berndt-Stra{\ss}e 3, D-64287 Darmstadt, Germany\\
$^{3}$Department of Materials Science, Chair of Materials Physics, Mountanuniversit{\"a}t Leoben, Jahnstra{\ss}e 12, A-8700 Leoben, Austria\\
$^{4}$Laboratory for Microstructures, Institute of Materials, Shanghai University, 200444, Shanghai, China\\
$^{5}$European Synchrotron Radiation Facility (ESRF), 38042 Grenoble, France \\
$^{6}$Laboratory of Metal Physics and Technology, Department of Materials, ETH Zurich, 8093 Zurich, Switzerland}

\begin{abstract}
Residual stress engineering is very widely used in the design of new advanced lightweight materials.  
For metallic glasses the attention has been on structural changes and rejuvenation processes.
High energy scanning X-ray diffraction strain mapping reveals large elastic fluctuations in metallic glasses after deformation under triaxial compression. 
Microindentation hardness mapping hints to a competing hardening-softening mechanism after compression and further reveals the complementary effects of stress and structure modulation. 
Transmission electron microscopy proves that structure modulation under room temperature deformation relates to the shear band formation that closely correlates to the distribution of elastic heterogeneities. 
Molecular dynamics simulations provide an atomistic understanding of the complex shear band activity in notched metallic glasses and the related fluctuations in the strain/stress heterogeneity. 
Thus, future focus should be given to stress engineering and elastic heterogeneity that together with structure modulation may allow to design metallic glasses with enhanced ductility and strain hardening ability.
\end{abstract}

\maketitle

%\pacs{}
\section{Introduction}

Given the metastable nature of metallic glasses (MGs) and the related rugged energy landscape with a large diversity of energy barriers makes the properties of this class of material to be easily tunable \cite{Isner2006,fan2017}. 
Nowadays there is a large number of methods available to alter the structure and corresponding properties of MGs \cite{saida2013recovery, kuchemann2018energy,wakeda2015controlled,pan2018extreme, park_2008, concustell2009structural, dmowski2010, 
ding2019ultrafast, feng2018rejuvenation}.  
Rejuvenation and relaxation are competing processes in controlling the energy of MGs towards higher and lower levels resulting, in the end, to a wide range of possible glassy states \cite{concustell2009structural, mahmoud2021identification}.
Rejuvenation of MGs is usually referred to increased disorder resulting in excess free volume and a raise in energy \cite{dmowski2010,Liu2010}. 
Nevertheless, rejuvenation is not always associated with structural softening and hardness reduction. 
Rejuvenation processes done by combining deformation protocols and thermal treatments revealed a density and hardness increase of some MGs  \citep{ge2017unusual, wang2017high, miyazaki2016prediction}.
The observed hardening behavior was associated to confined microplasticity and local accumulation  of  irreversible  compressive  strains \cite{Deng2012,Zhao2017}. 
Furthermore, deformation by cold-rolling and shot-peening results in a transition from softening to hardening behavior that was associated with the accumulation of compressive residual stresses overcompensating shear softening and the corresponding 
free volume generation \cite{Stolpe2014, zhang2006}. 
Highly rejuvenated MGs obtained under triaxial compression revealed both, strain  hardening  and enhanced ductility \cite{pan2018extreme,Pan2020}. 
The exceptionally efficient strain-hardening and suppression of shear-banding were associated with structural relaxation of the extremely rejuvenated MG \cite{Pan2020}.

While structural rejuvenation received much attention, there has been less interest in recent years in exploring stress engineering in MGs.
Due to their very low thermal conductivities and the high cooling rates involved in the quenching process, MGs undergo differential cooling and large thermal gradients are generated across the sample leading to residual stress buildup \cite{Ustundag1998,Wang2011}.
Residual stresses and stress field gradients affect shear band dynamics and the overall plastic deformation \cite{Kosiba2020,zhang2006}, hardness \cite{Wang2011,Haag2014} and even magnetic properties of MGs \cite{Wang2012a}.   
Recently, molecular dynamics (MD) simulations of stress modulated MG heterostructures revealed that strain hardening together with tensile ductility can be attained by only modulating the
internal residual stress without altering the structure \cite{Yuan2021}. 

Here, we focus our attention on clarifying the key factors underlying the rejuvenation process under triaxial compression and distinguish between structural and elastic fluctuations.
Using a similar setup proposed as by Pan et al. \cite{Pan2020} at the ESRF Extremely Brilliant Source, strain mapping of extremely deformed MG samples was performed. 
This novel approach allows to distinguish hydrostatic from deviatoric strains and to quantify structural and stress heterogeneities  \cite{ShakurShahabi2016,scudino2019}.
Transmission electron microscopy (TEM) is also applied to highlight intense shear band activity during the room temperature deformation.
Hardness mapping with microindentation is used to emphasize the competing effects of stress and structure modulation.
Finally, an atomistic understanding of the confined deformation mechanisms in a notched MG is provided by MD simulations.

\begin{figure*}[t]
\centering
\includegraphics[angle=0,width=0.85\textwidth]{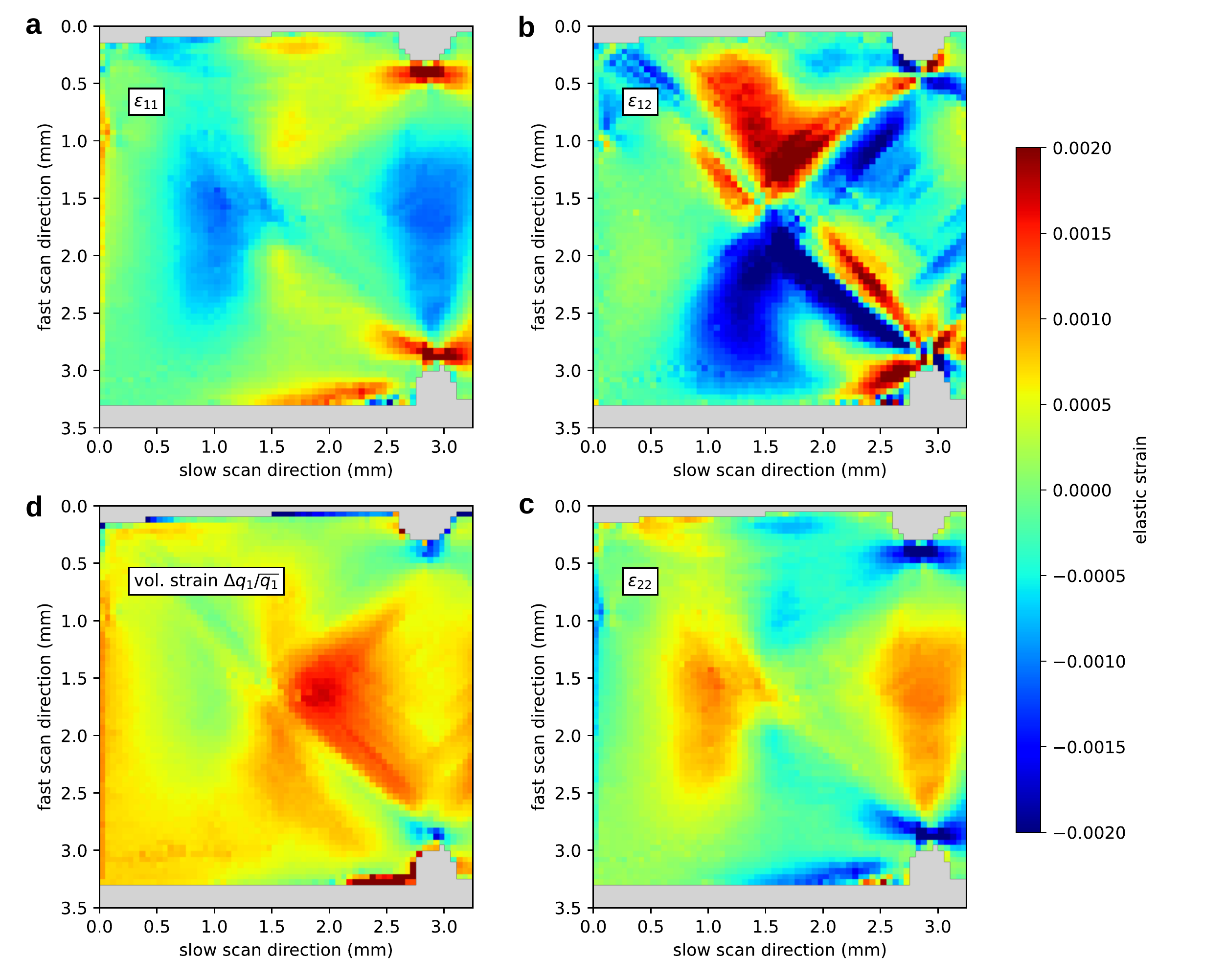}   
\caption{Plane strain distribution in the Zr$_{61}$Ti$_2$Cu$_{25}$Al$_{12}$ MG deformed under triaxial compression. 
The deviatoric plane strains $\epsilon_{11}$ (a), $\epsilon_{12}$ (b), and  $\epsilon_{22}$ (c) determined from the ellipticity of the first amorphous diffraction maximum. 
(d) The volumetric strain $\Delta q / \Delta q_0$ is represented to show the change in free volume.}
\label{1}
\end{figure*} 

\begin{figure*}[t]
\centering
\includegraphics[angle=0,width=0.99\textwidth]{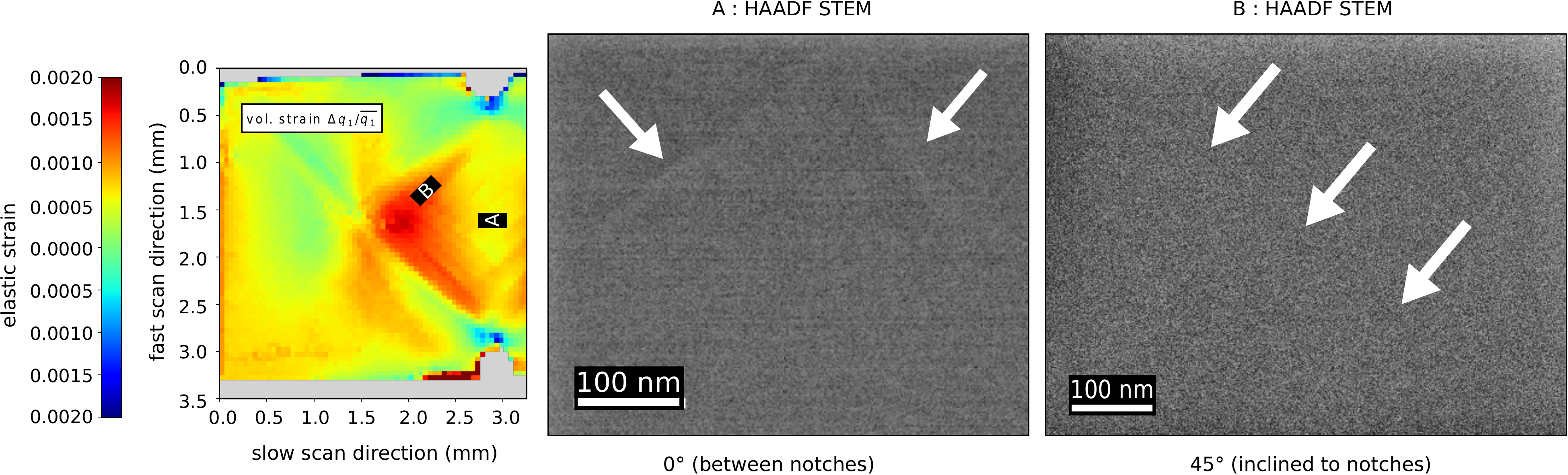}   
\caption {HAADF-STEM images of two random selected-areas.
Two HAADF-STEM images were taken from lamellae with thickness of $\approx$ 100 nm between the notch area and along the shear front (black rectangles marked with A and B in the volumetric strain map).
For each cross-section HAADF-STEM images show alternating contrast changes, which are related to presence of shear bands in the material. 
Arrows pointing at the shear bands are a guidance for the eye.}
\label{2}
\end{figure*} 

\begin{figure*}[t]
\centering
\includegraphics[angle=0,width=0.85\textwidth]{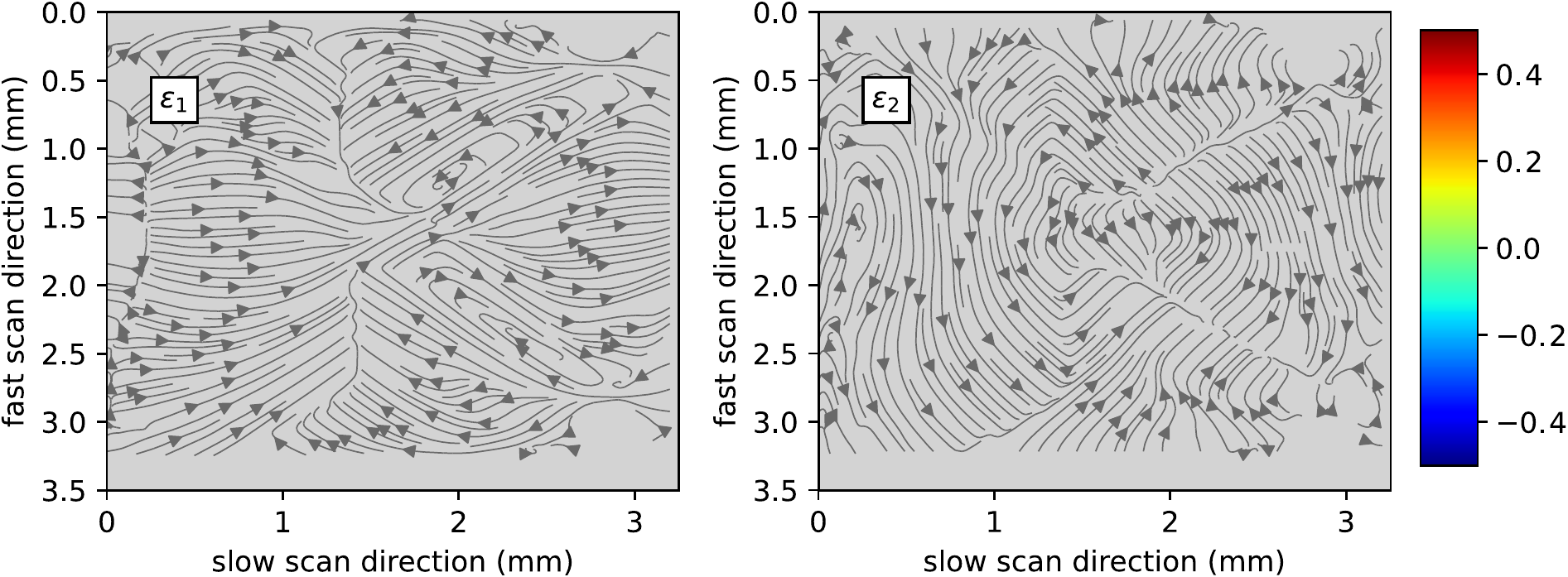}   
\caption{Eigenvalue directions and values determined from the deviatoric strains. 
The flow lines for the  strain directions $\epsilon_1$ and $\epsilon_2$ reveal turbulent behavior around the notches and at along the shear fronts due to the change from positive to negative strains.}
\label{3}
\end{figure*} 

\section{Results}

The plane strain distribution can be experimentally determined with high spatial resolution in MGs \cite{binkowski2015,ShakurShahabi2016,scudino2019}.
In Fig. \ref{1}  we show for the first time the antisymmetric nature of the strain evolution in the MG sample at the notch with unprecedented resolution.
As shown in Figs. \ref{1}a and c, the elastic strains parallel and perpendicular to the loading direction ($\epsilon_{11}$ and $\epsilon_{22}$) display a highly heterogeneous profile that seems to be stabilized even after 
releasing the external load.
This behavior indicates a highly inhomogeneous deformation mechanism, i.e., shear banding \cite{ShakurShahabi2016,scudino2019,Kosiba2020} and, hence, it further confirms that, at low temperature, the shear bands are the only plastic carriers in MGs.
HAADF-STEM images of random selected-areas along the shear fronts (of highest volumetric strain) and between the notch area highlight shear band activity (see Fig. \ref{2}).

The $\epsilon_{11}$ component shows localization of dilatation at the notches and along the two shear fronts that extend even away from the notches.
Contraction can be observed between the notch area and ahead of the intersection point of the two shear fronts. 
The $\epsilon_{22}$ component shows exactly an inverse trend to $\epsilon_{11}$.
The strong variations in the shear strain ($\epsilon_{12}$) further highlight the presence of a complex interpenetrating network of shear bands (Fig. \ref{1}b) \cite{ShakurShahabi2016}.
Variations in the elastic strain, from compressive to tensile states, are not only present across the two shear fronts but can be also seen between the notches.   
The latter Eigenvector evaluation also allows to determine volumetric strains (free volume) that are in good agreement with the reciprocal space dilatation evaluations (Fig. \ref{1}d).
The highest volumetric strain can be seen at the intersection point of the two shear fronts, where a high density of shear bands concentrates, and shows a descending gradient towards the two notches.

The variations in the Eigenvectors of the strain tensor indicates the direction of the principal strain axes and allows for the localization of shear bands in MGs \cite{scudino2018}. 
The Eigenvectors exhibit an antisymmetric sigmoidal profile consisting of large angular variation at the position of SBs.
The evaluation of the Eigenvectors in Fig. \ref{3} highlights a turbulent strain distribution with a localization of these turbulent areas close to the notches and to the area of highest dilatation.
Such a turbulent strain distribution correlates to the change in the sign of the strain from compressive to tensile strain. 
The sigmoidal profile is stronger at the position of the two main shear fronts but weaker variations in the sign of the strain can be also observed between the notches highlighting the presence of shear activity.  

\begin{figure}[t]
\centering
\includegraphics[angle=0,width=0.85\textwidth]{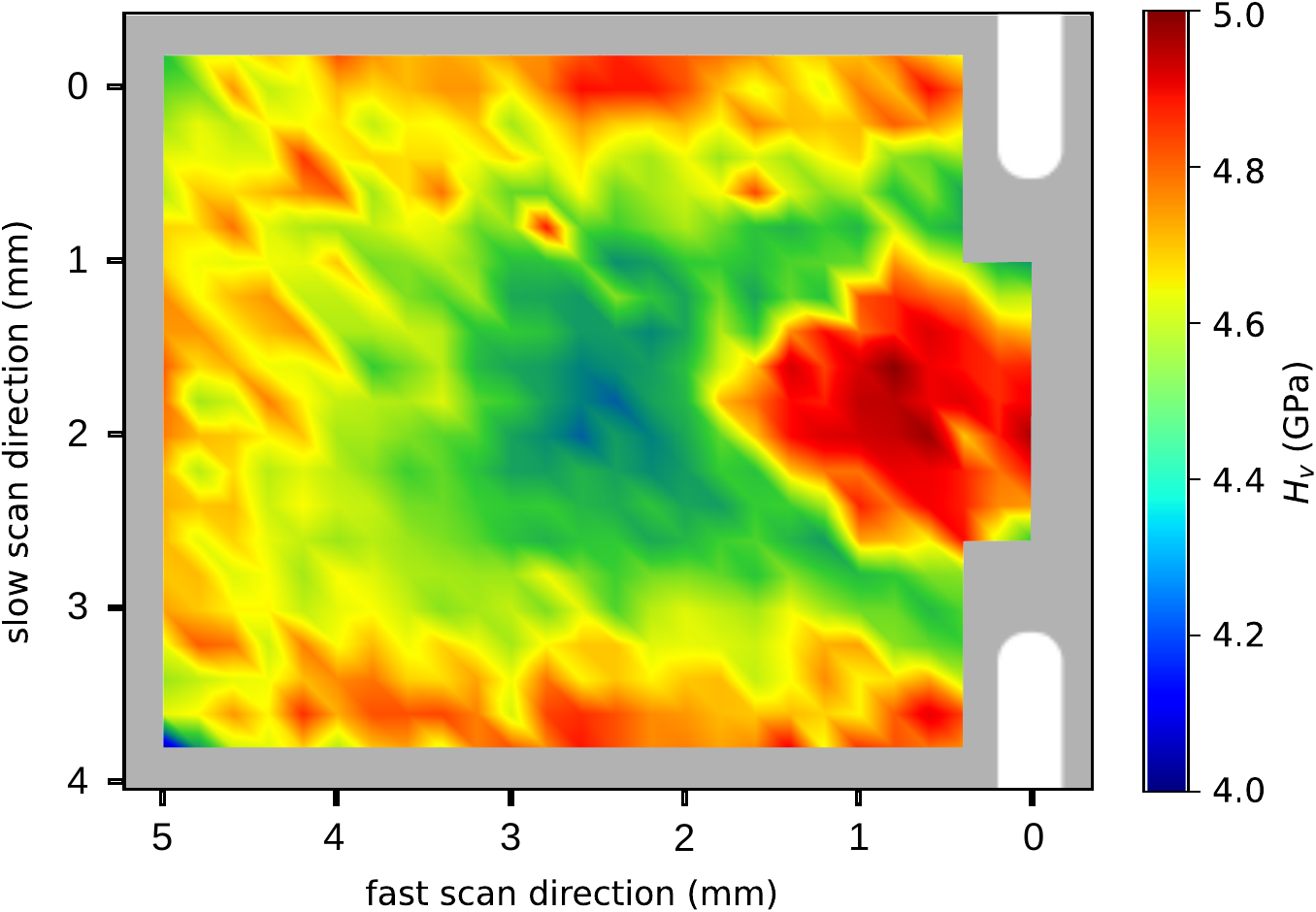}   
\caption{Hardness mapping with microindentation.
The hardness map indicates a high local heterogeneity over the sample and hints at an unsymmetrical hardening/softening after compression.}
\label{4} 
\end{figure}

To distinguish between structural and stress fluctuations in the extremely deformed MG sample, hardness mapping over half of the sample was performed with microindentation.
The hardness map indicates a pronounced local heterogeneity across the sample (see Fig. \ref{4}).
As expected, the hardness decreases monotonically with increasing free volume \cite{Yoo2009}.
This correlates to the increase in the volumetric strain, especially at the intersection point of the two shear fronts of highest dilatation 
(highlighted in Fig. \ref{1}d).
However, below the two shear fronts where the volumetric strain is still higher, as compared to the undeformed sample, the hardness shows the highest values. 
The shear induced softening and the related large volumetric strain do not decrease the hardness as we would have expected.  
The only reasonable explanation for such a questionable hardness mapping could be the presence of residual stresses and stress field gradients that overcompensate the shear softening.   
Such a behavior could already predict that besides the structural fluctuations (free volume) also the deviatoric strains (stress heterogeneities) may play a major role in the hardness behavior and in general in the mechanical properties of MGs \cite{zhang2006}.

\begin{figure}[t]
\centering
\includegraphics[angle=0,width=0.85\textwidth]{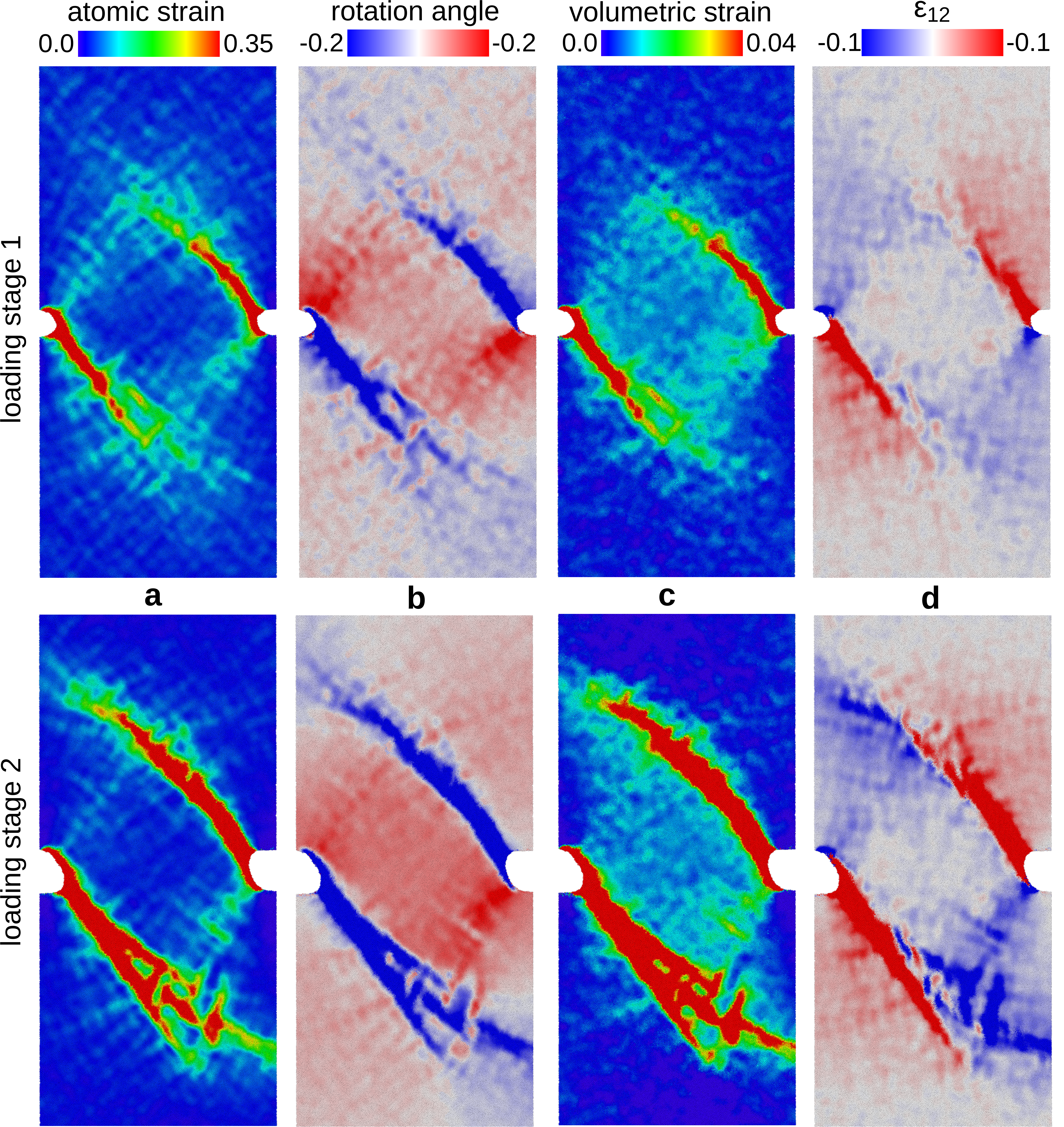}   
\caption{MD simulations of the deformation mechanisms in the notched MG alloy.
Structural variations in the simulated Cu$_{64}$Zr$_{36}$ MG at two loading stages, at an early stage 1 of deformation of shear band formation (upper panels) and a later stage 2 when the shear bands interact (lower panels). 
Evolution of the atomic von Mises strain (a), the rotation angle (b), the volumetric strain (c) and the strain tensor ($\epsilon_{12}$) (d).}
\label{5}
\end{figure}

MD simulations were used to provide an atomistic understanding of the complex deformation mechanisms between the two notches.
A simplified rectangular sample with two notches was deformed.
When the yield commences four shear fronts initiate from the two notches (as highlighted by the atomic strain in Fig. \ref{5}a, upper panel).
During loading only two shear bands become dominant and define the shear direction.
Even with continued deformation of the sample, these two shear bands do not become critical and do not transect the entire sample (see the lower panel in Fig. \ref{5}a).
Although the other two shear bands do not mature, their strain fields perturb the percolation and further propagation of the dominant shear bands.  
A better view of the shear band interaction mechanism is given by the sign of the rotation angle, generally used to identify vortex-like (rotating) units in glassy structures (Fig. \ref{5}b) \cite{Sopu2020,Sopu2021}.
The sign of rotation defines the shear direction and has high implication on the dynamics of shear band interaction.
When two shear bands intersect, vortexes of opposite rotation directions come into contact and the further movement of the shear bands is hindered \cite{Sopu2021}. 
This process blocks the initial shear front and changes the shear band dynamics and morphology, ultimately leading to shear band blocking, branching and multiplication \cite{Sopu2020}.
Shear band multiplication and blocking processes are particularly important as they play a decisive role in enhancing the plasticity of MGs, what could explain the extended compressive plastic flow in the central notched region \cite{pan2018extreme}.

However, structural variation is not only confined to the narrow shear bands.
When analyzing the volumetric strain one can see that the local density does not increase only at the position of the shear bands (Fig. \ref{5}c). 
The entire area at the notch, between the four shear fronts, shows larger variations with respect to the structure away from the notches.
The complex shear band network confines the material and changes the local strain/stress state. 
In fact, this could be responsible for the observed homogeneous rejuvenation of MG when deformed under triaxial compression \cite{Pan2020}.  
Indeed, when analyzing the strain tensor ($\epsilon_{12}$) distribution obtained from experiments (Fig. 1b) and MD simulations (Fig. \ref{5}d) a striking resemblance can be observed.
The variation in the elastic strain from compressive to tensile states closely follows the shear band front.  
Moreover, at the position where the  shear band branches, the sign of the strain shows a sinusoidal strain profile (tensile vs. compressive), highlighting the formation of multiple shear bands. 

\section{Discussion}

From the above results it can be stated that triaxial deformation of MGs induces both plane (directional) and deviatoric (volumetric) strains.
Hence, the reason for the observed extreme rejuvenation corresponds to both, structural softening and residual stresses imparted during the deformation process.   
Additionally, the stress heterogeneity is strongly modulated and correlated with the intense shear band activity, as indicated by the shear fluctuations (change in the sign from positive to negative) in the area between the two notches (Fig. \ref{1}).
It is well accepted that shear banding is not only confined to a narrow thin layer of about 20~nm but extends over a longer range (tens of micrometers) across the shear band  \cite{Maass2011,liu2018,scudino2018}.
Experimental observations of high-energy XRD across an individual shear band revealed that the strain at opposite sides of a shear band changes sign from compressive to tensile \cite{scudino2019,scudino2018,Kosiba2020}.
Likewise, the strains/stresses are created in an inhomogeneous manner during triaxial compression (see Fig. \ref{1}).
Moreover, these fluctuations are even there after unloading, cutting and polishing the highly deformed sample. 
Previously it has been stated that residual stresses are released after the small sample is machined within the notches of a pre-deformed rod \cite{Pan2020}.
There, the degree of rejuvenation was quantified using the exothermic heat of relaxation ($\Delta H_{rel}$).
The measured $\Delta H_{rel}$ was mainly related to structural changes, and the volume accumulation, and a only small fraction of this energy corresponds to residual stresses \cite{Stolpe2014}.
Consequently, the contributions of stress heterogeneity/fluctuations and residual stresses to rejuvenation and strain-hardening were underestimated \cite{Pan2020}.
Nevertheless, the presence of residual stresses is reflected in the observed large change in density that exceeds 0.2$\%$ (Fig. \ref{1}) \cite{johnson2005,wang_2006}.
Indeed, the relative change in density as a function of cooling rate is usually lower than 0.07 $\%$ \cite{Hu2001,Fu2013}.
Similar variations in density were found in cold-rolled metallic glasses \cite{Haruyama2013}.
Large variations in density up to 0.16 $\%$ have been observed in heavily deformed MGs \cite{Vempati2012}.
In this case, the density fluctuations correspond to both, enhanced disorder (structural rejuvenation) and the stress constraint introduced by the residual stresses, respectively.

The effects of compressive and tensile residual stresses after mechanical loading are especially important when considering that they can overcompensate or undercompensate the softening caused by shear banding and free volume generation \cite{Launey2008,Yuan2021}.
These competing effects may help to explain why the hardness map does not perfectly correlate to the volumetric strain (free volume), especially at the position between the two notches (see Fig.~\ref{1}c).
Here, the volumetric strain shows higher values (Fig. \ref{1}c) and hence, lower hardness values are expected. 
Contrarily, the microhardnesses map indicates hardening in the region between the notches (Fig. \ref{4}). 
In this area the elastic strains $\epsilon_{11}$ and $\epsilon_{22}$ display large fluctuations (Figs. \ref{1}a, b and d). 
Since the sample is symmetric perpendicular to the loading direction also $\epsilon_{33}$ is tensile.
The residual tensile stress can be seen as a back stress \cite{Zhu2019, Yuan2021} that must be overcome by the indent to penetrate the system and to switch locally to a compressive stress state.
Namely, the contact pressure underneath the indent that is required to initiate the yield significantly increases in these regions \cite{Wang2011,Chen2008}.

\section{Conclusion}
In conclusion, this research has significantly advanced our understanding of MG rejuvenation under constrained deformation protocols.
Based on high energy scanning X-ray diffraction strain mapping, STEM and microindentation one can distinguish between structural and elastic fluctuations, the two key factors for the observed extreme rejuvenation in triaxial compression.
STEM characterization shows that structural rejuvenation under room temperature deformation relates to shear-induced softening and dilatation (large volumetric strain). 
An interpenetrating network of multiple shear bands not only increases the free volume but also results in large variations in the elastic strain. 
High energy scanning X-ray diffraction strain mapping reveals large elastic fluctuations corresponding to the intense shear band activity.
The complex superimposed spatial stress and structure heterogeneities reflect in the weak correlation between hardness and structural softening and support the competing hardening-softening mechanism.
Hence, the classical description of rejuvenation should be reconsidered and future focus should be also given to stress-driven modulation.
Stress and structural engineering can be used simultaneously to enhance rejuvenation beyond the limits known so far and, consequently, can help to design MGs with enhanced ductility and strain hardening ability.

\section{Acknowledgments}
The authors acknowledge financial support by the Deutsche Forschungsgemeinschaft (DFG) through Grant No. SO 1518/1-1, the European Research Council under the ERC Advanced Grant INTELHYB (grant ERC-2013-ADG-340025), Austrian Science Fund
(FWF): Y1236-N37 and the China Scholarship Council (CSC, 201806220096).
We acknowledge the European Synchrotron Radiation Facility (ESRF) for provision of synchrotron radiation facilities and we would like to thank Dr. Carlotta Giacobbe  for assistance and support in using beamline ID 11.
Calculations for this research were conducted on the Lichtenberg high performance computer of the Technische Universit\"at Darmstadt.

\section{Methods}
\subsection{Preparation of metallic glass samples}
\vspace{-0.5\baselineskip}
\hfill\\
Metallic glass with a nominal composition of Zr$_{61}$Ti$_2$Cu$_{25}$Al$_{12}$ (at.$\%$) was prepared by arc-melting a mixture of high-purity ($>$ 99.9$\%$) metals at least four times in a titanium-gettered high-purity argon atmosphere. 
Cylindrical samples with a diameter of 4.5 mm and a length of 70 mm were fabricated by suction casting into a copper mold. 
The structure of the glassy phase of the as-cast MG samples was ascertained by X-ray diffraction (XRD) using a Rigaku Dmax-2550 diffractometer with Cu-K$\alpha$ radiation generated at 40 kV. 
Specimens with a length/diameter ratio of 2 were cut from the rods by a diamond saw, and the surface roughness of the two parallel ends were carefully polished. 
Notched specimens with a circumferential notch dimension (notch diameter, d $\times$ notch height, h) of 2.5 $\times$ 0.5 mm$^2$ were produced using a 0.5 mm diameter round-head grooving knife at a feed rate of 0.06 mm/r in a lathe machine with a running speed of 
500 r/min followed by fine polishing and final cleaning in an ultrasonic bath. 
Then, the notched specimens were compressed along the cylindrical axis at room temperature using MTS CMT5205 device at a strain rate of 10$^{-4}$s$^{-1}$ until the width of the notch was reduced by $\approx$ 40$\%$ plastic flow.
Due to the notch constrain the central zone is deformed under triaxial compression, suppressing relaxation \cite{Pan2020}.
The deformed specimens were then machined along the central lateral and longitudinal planes. 
The longitudinal cross-sections were mechanically polished to mirror finish for the following high energy XRD and nanoindentation experiments.

\subsection{Structural characterization}
\vspace{-0.5\baselineskip}
\hfill\\
High energy scanning XRD was used to perform strain mapping on the cross-section of the deformed specimen, which was carried out at the nanostation of the beamline ID11 at ESRF.
A wavelength of 0.01919~$nm$ was used and the beam-size was focused to 500 $\times$ 500~$nm$. 
The region of interest was scanned with 3500 points with a spacing of 1 $\mu m$ each in the fast scanning direction and 65 with a spacing of 50~$\mu m$ in the slow scanning direction leading to a total of 227500 two-dimensional diffraction patterns 
recorded with an exposure time of 10~$ms$ with an Eiger CdTe 4M fast pixel array detector. 
In the fast direction 50 frames were summed to increase the signal-to-noise ratio. 
The newly upgraded Extremely Brilliant Source (EBS) of ESRF allowed measurement of half of the sample in about 2 hours. 

The data was calibrated using a CeO$_2$ reference and integrated using the pyFAI software. 
For the evaluation of plane strains, the integration was performed in 72 azimuthal sections. 
The volumetric strains were evaluated using the average center of mass of the first strong diffraction maximum $q_{COM}/q_0$ of the glassy material. 
The fitting procedure is reproduced for a representative diffraction pattern in Fig. S1.

For the plane strains $\epsilon_{11}$, $\epsilon_{12}$ and $\epsilon_{12}$, the normalized center of mass of the first diffraction maximum $q(\phi)/q_{COM}$ was fitted using the following quation for the  azimuthal angle $\phi$ similar 
to the established procedure \cite{ShakurShahabi2016}
\begin{equation}
\phi = \epsilon_{11} cos (\phi)^2 + \epsilon_{12} \mathrm{cos} (\phi)  sin(\phi) + \epsilon_{22} sin(\phi)^2.
\end{equation}

The high flux of EBS and new fast pixel array detectors with single photon counting ability allow correlation of the mesoscale information occurring in the plastic deformation of the notched MG sample with the formation of individual shear bands
and their interactions. 
Strain mapping  allows identification and visualization of shear bands in the bulk material in 2D.

Thin lamella for scanning transmission electron microscopy (STEM) were prepared with a final thickness of 100~$nm$ by focused ion-beam milling using a Zeiss Auriga dual-beam workstation. 
One lamella was taken from the middle part between the two notches of the specimen. A second lamella was taken from a position inclined by 45$^{\circ}$ with respect to the notch direction. 
High Angle Annular Dark Field (HAADF) STEM images were recorded for each lamella with a spot size of 1~$nm$. 

Macroscopic Vickers hardness maps were recorded using a universal fully-automatic DuraScan70G5 laboratory ZwickRoell hardness tester. 
The areal distribution of indentation points was recorded in a grid of 200~$\mu m$ x 200~$\mu m$ with a total number of 522 indents in a matrix of 26x21. 
To avoid the influence of the elastic-plastic stress field, the indents were spaced three diagonals apart. 
The indentation depth of each indent was approximately 15~$\mu m$, calculated using the 136$^{\circ}$ opening angle of the pyramid-shaped diamond indenter. 
A load of 9,807~$N$ was applied for each impression, which corresponds to the small load range of HV1. 

\subsection{Molecular dynamics simulations}
\vspace{-0.5\baselineskip}
\hfill\\
A qualitative and elemental atomistic picture of the confined deformation in a notched MG was provided by means of classical molecular dynamics simulations carried out with the  LAMMPS software \cite{plimpton1995}.
Because no appropriate interatomic potentials exist for Zr$_{61}$Ti$_2$Cu$_{25}$Al$_{12}$, the Cu$_{64}$Zr$_{36}$ MG was used as a prototype material since a reliable EAM potential was made available by Mendelev et al. \cite{Mendelev2007}.
The amorphous Cu$_{64}$Zr$_{36}$ alloy structures predicted by the potential are in good agreement with X-ray diffraction data and the potential can be reliably used to simulate the structure and properties of amorphous Cu-Zr alloys \cite{Sopu2020,Sopu2021}.

The starting liquid structure was created by randomly distributing 968000 atoms in a box of $111.0 \times 55.5 \times 2.5\;nm$ with periodic boundary conditions. 
The amorphous structure was obtained by quenching from the melt starting from 2000 K.
More details about quenching procedure can be found in ref. \cite{Moitzi2020}.
The cooled sample was replicated twice along the $z$-direction and two notches were created along the $y$-direction.
Open boundaries in $y$-direction and periodic boundaries in the $x$- and $z$-directions were applied.
The time step was set to 2 fs.

Uniaxial tensile simulations along the $z$-direction were performed on the notched MG sample under a constant engineering strain rate of 4$\times$10$^7~s^{-1}$.
The stress along $x$-direction was relaxed to allow for lateral contraction using Nose/Hoover barostat \cite{plimpton1995}.
A low temperature of 50~K was chosen to compensate for the very high strain rate and guarantee a regime of localized deformation and shear band formation. 
The deformation mechanisms and the related structural fluctuations at two different strain levels were analyzed using the OVITO software \cite{Stukowski2010} by calculating and visualizing the von Mises strains, simultaneously occurring directional rotation fields, 
volumetric strains and the strain tensors. 
All these micromechanical quantities were evaluated for each atom in the system from the relative motion of the neighboring atoms within a cutoff range of 1~$nm$.

\normalem
%\bibliographystyle{unsrtnat}
%\bibliography{References}

\end{document}